\documentstyle[aps,epsf,twocolumn]{revtex}
\def\la{\langle} 
\def\ra{\rangle} 
\def\be{\begin{eqnarray}} 
\def\ee{\end{eqnarray}}

\newcommand{\eq}{\begin{equation}} \newcommand{\eqx}{\end{equation}}

\newcommand{\eqn}{\begin{eqnarray}} \newcommand{\eqnx}{\end{eqnarray}}
\newcommand{\f}[2]{\frac{#1}{#2}}

\newcommand{\arr}[4]{
\left(\begin{array}{cc}
#1&#2\\
#3&#4
\end{array}\right)}

\begin{document}
\draft

\title{\bf Two-Color QCD and Aharonov-Bohm Fluxes}

\author{ {\bf Romuald A. Janik}$^1$, {\bf Maciej A.  Nowak}$^{1,2}$ ,
{\bf G\'{a}bor Papp}$^{3}$ and {\bf Ismail Zahed}$^4$}

\address{$^1$ Department of Physics, Jagellonian University, 30-059
Krakow, Poland.
\\ $^2$ GSI, Planckstr. 1, D-64291 Darmstadt, Germany
\\ $^3$ITP, Univ. Heidelberg, Philosophenweg 19, D-69120 Heidelberg, 
	Germany \& \\ Institute for
Theoretical Physics, E\"{o}tv\"{o}s University, Budapest, Hungary\\
$^4$Department of Physics and Astronomy, SUNY, Stony Brook, 
New York 11794, USA.}
\date{\today} \maketitle

\begin{abstract}
We investigate the effects of several Abelian Aharonov-Bohm fluxes $\phi$ on 
the Euclidean Dirac spectrum of light quarks in QCD with two colors. A 
quantitative change in the quark return probability is caused by the fluxes, 
resulting into a change of the spectral correlations.
These changes are controlled by a universal function of $\sigma_L \phi^2$ 
where $\sigma_L$ is the pertinent Ohmic conductance. The quark return 
probability is sensitive to Abelian flux-disorder but not to $Z_2$ 
flux-disorder in the ergodic and diffusive regime, and may be used as 
a probe for the nature of the confining fields in the QCD vacuum.

\end{abstract}

{\bf 1.}
In a finite Euclidean volume $V$ light quarks exhibit properties
analogous to that of electrons in small metallic grains~\cite{USPRL}. 
The finite volume allows for a discrimination of length scales with the 
emergence of a diffusive regime characterized by the diffusion 
constant $D\approx 0.22$ fm~\cite{USPRL}. Although QCD confines, 
the light quarks diffuse in four dimensions as a result of the 
spontaneous breaking of chiral symmetry~\cite{USPRL,BOOK}.

In two-color and two-flavor QCD the quarks are in the pseudoreal
representation 
of the flavor group~\cite{PESKIN,LEUTSMILGA}, and the spontaneous breaking
of chiral symmetry is accompanied by the occurrence of five Goldstone
modes: 3 pions and a baryon and an antibaryon. The five Goldstone modes
decay weakly, with a mass that satisfies the Gell-Mann--Oakes--Renner 
(GOR) relation~\cite{PESKIN}. The latter mass vanishes in the chiral limit.

The nature of the Goldstone modes and the characteristics of the light
quark spectrum are intimately related in the general context of chiral
disorder~\cite{USPRL}. In particular, two-color QCD is characterized by
diffuson (diagonal, pionic) and cooperon (interference, baryonic)
contributions in
the semi-classical analysis. The cooperons correspond to weak-localization 
(coherent back-scattering) contributions to the quark return 
probability~\cite{USPRL,ALTSHULLER}. They
are soft because the baryons in two-color QCD are diquarks and soft. 
QCD with three colors admit only diffusons as the diquarks
are expected to be heavy. As well known in disordered metals~\cite{MONTAM}, 
the weak-localization contribution is altered by the presence of an external 
parameter that breaks time-reversal symmetry. An example is an Aharonov-Bohm 
flux.

In this letter we consider the effects of several Abelian Aharonov-Bohm 
fluxes on the light quarks in a finite Euclidean volume with two-color QCD.
In many ways, our analysis in two-color QCD will parallel recent analyses
of electrons in disordered metals~\cite{MONTAM}.  We discuss these effects 
on the light quark return probability, and analyze their importance on the 
spectral correlations.  The fluxes cause a periodic interpolation in
the spectral statistics between orthogonal and unitary ensembles. 
We show that the quark return probability is sensitive to Abelian 
flux-disorder but not to $Z_2$ disorder, and discuss the relevance of
this result for the nature of the QCD vacuum.

\vskip 0.3cm

{\bf 2.}
In a finite Euclidean volume $V=L^4$ pierced by Abelian 
(essentially electromagnetic) fluxes 
$\phi_{\mu}=(\phi_1,\phi_2,\phi_3,\phi_4)$,
the light quarks in the background gluon field $A$ satisfy the following
Dirac equation
\be
i\nabla \!\!\!\!/ [A] \,q_k =\lambda_k [A, \phi] \, q_k \,\,.
\label{01}
\ee
subject to the boundary condition
\be
q_k (x+L_{\mu}) =-e^{i2\pi \phi_{\mu}}\,\,q_k (x) \,.
\label{02}
\ee
The $\phi$'s are given in units of an electromagnetic 
flux quantum $2h/e$~\cite{NOTE2} set to 1 for
convenience.
Through (\ref{02}) the quark spectrum (\ref{01}) depends explicitly on $\phi$.
We note that the Abelian fluxes through (\ref{02}) are simple holonomies
along each of the four directions of the four-volume: 
$e^{i2\pi x\cdot\phi/L}$. The eigenvalue problem (\ref{01}-\ref{02})
is equivalent to
\be
\left( i\nabla \!\!\!\!/[A]  + \frac{2\pi}L \,\gamma\cdot\phi\right)
\,q_k =\lambda_k [A, \phi] \, q_k \,\,.
\label{001}
\ee
with anti-periodic boundary conditions.

The probability $p(t, \phi)$ for a light quark to start at $x(0)$ in $V$ and 
return  back to the same position $x(t)$ after a proper time duration $t$, is
\be
p(t, \phi )= \frac {V^2}N
\Big\la |\la x(0)|e^{i(i\nabla \!\!\!\!/[A] +im)|t|}|x(0)\ra|^2\Big\ra_A\,.
\label{1}
\ee
The averaging in (\ref{1}) is over all gluon 
configurations using the unquenched two-color QCD measure. The
normalization in (\ref{1}) is per state, where $N$ is the total number
of quark states in the four-volume $V$. Equation~(\ref{1}) may be 
written in terms
of the standard Euclidean propagators for the quark field,
\be
p(t, \phi ) = \frac {V^2}N &&\lim_{y\to x}
\int \frac {d\lambda_1d\lambda_2}{(2\pi)^2} 
\,e^{-i(\lambda_1-\lambda_2) |t|}\nonumber\\&&
\times\Big\la {\rm Tr}\left( S(x,y;z_1,\phi) 
S^{\dagger} (x,y; z_2,\phi)\right)\Big\ra_A
\label{des2}
\ee
generalizing to $\phi \neq 0$ the results in~\cite{USPRL}.
Here  $z_{1,2}=m-i\lambda_{1,2}$, and
\be
S(x,y; z, \phi) = \la x| \frac 1{i\nabla \!\!\!\!/[A] + iz} |y\ra \,.
\label{des3}
\ee
subject to the boundary condition
\be
S(x+L_{\mu}, y;z,\phi) =-e^{i2\pi\phi_{\mu}}\,\,
S(x,y;z,\phi) \,.
\label{bound}
\ee
Setting $\lambda_{1,2}=\Lambda\pm \lambda/2$ and neglecting the effects
of $\Lambda$ in the averaging in (\ref{des2}) (zero virtuality),
the correlation function in (\ref{des2}) relates (in general) to 
an analytically continued pseudoscalar correlation function, as 
the eigenvalues $q_k$ and $\gamma_5 q_k$ are 
pair-degenerate (chiral)~\cite{USPRL}. 

For two-color QCD there is an extra symmetry~\cite{PESKIN,LEUTSMILGA}
that makes the pseudoscalar correlation function degenerate with certain
diquark correlation functions in the flux-free case. Indeed, for $\phi=0$
and ${\bf K}=-T^2C{K}$ the eigenvalues $q_k$ and 
${\bf K}\, q_k$  are  also
pair-degenerate. Here $C$ is the charge-conjugation matrix, 
$T^2$ the color matrix and $K$ the (right-left) complex-conjugation. 
For instance, for two flavors the three pions are degenerate with the 
spin-zero baryon and antibaryon. 

In general, we may rewrite (\ref{des2}) in the form
\be
p(t, \phi ) = \frac {EV^2}{2\pi N} \lim_{y\to x}{}
\int \frac {d\lambda}{2\pi} 
\,e^{-i\lambda |t|} {\bf C}_{G} (x,y; z, \phi)
\label{des55}
\ee
with $z=m-i\lambda/2$ and $E=\int d\Lambda$. For $\phi=0$, the 
above symmetries allow us to write (two flavors)
\be
\label{des5}
&&{\bf C}_{G} (x,y;z, 0 ) = \\
&&+\frac 12\Big\la {\rm Tr}\left(
S(x, y;z, 0 ) i\gamma_5\tau^2 S(y,x;z, 0 ) i\gamma_5\tau^2\right)
\Big\ra_A \nonumber\\
&&+\frac 12\Big\la {\rm Tr}\left(
S(x, y;z, 0 ) \loarrow{\bf K}i\gamma_5\tau^2 S(y,x;z, 0 ) 
\roarrow{\bf K}i\gamma_5\tau^2\right)
\Big\ra_A \nonumber
\ee
The flavor matrix $\tau^2$ was introduced 
to retain  the connected parts in the correlator version~\cite{NOTE5}. 
The first contribution is pionic (diffuson), while the second one is baryonic 
(cooperon). Both are bosonic and identical. The decomposition (\ref{des5}) is
commensurate with the semi-classical description (long paths) where
the time-reversed trajectories are retained~\cite{USPRL,MONTAM}. 
For $\phi\neq 0$, the first 
symmetry (chiral) is retained while the second one is upset. Indeed, now the 
contributions in (\ref{des5}) are no longer the same as the fluxes add in the 
cooperon, but cancel in the diffuson. For long paths, we have
\be
{\bf C}_{G} (x,y;m, \phi ) \approx &&+\frac 1{2V} \sum_Q e^{iQ\cdot (x-y)} 
\frac {\Sigma^2}{F^2} \frac 1{Q^2+m_{G}^2}\nonumber\\
&&+\frac 1{2V} \sum_Q e^{iQ\cdot (x-y)} 
\frac {\Sigma^2}{F^2} \frac 1{\tilde{Q}^2+m_{G}^2}
\label{des06}
\ee
with $Q_{\mu} =n_{\mu}2\pi/L$ and 
$\tilde{Q}_{\mu}=(n_{\mu}+2\phi_{\mu}) 2\pi/L$.
Here, $\Sigma=|\la \overline{q} q\ra |$ and $F$ is the weak decay
constant for the Goldstone modes of mass $m_G$, each of which are 
flux independent  to leading order.

Using the GOR relation $F^2m_{G}^2=m\Sigma$
and the analytical continuation $m\rightarrow m-i\lambda/2$, we 
find
\be
{\bf C}_{G} (x,y;z, \phi) \approx &&+\frac 1{2V} \sum_Q e^{iQ\cdot (x\!-\!y)} 
\frac {2\Sigma}{-i\lambda\!+\!2m\!+\!DQ^2}\nonumber\\
&&+ \frac 1{2V} \sum_Q e^{iQ\cdot (x\!-\!y)} 
\frac {2\Sigma}{-i\lambda\!+\!2m\!+\!D\tilde{Q}^2}
\label{des6}
\ee
with the diffusion constant $D=2F^2/\Sigma$. 
Inserting (\ref{des6}) into (\ref{des55}), and noting that
$E/\Delta=N$ and $\rho=1/\Delta V$,
with $\Sigma=\pi \rho$, we conclude after
a contour integration that
\be
p(t, \phi) = \frac 12 e^{-2m|t|}\sum_Q \left( e^{-DQ^2 |t|} +
e^{-D\tilde{Q}^2 |t|}\right)\,.
\label{des7}
\ee
The cooperon contribution is periodic in the flux $\phi$
with periodicity $\phi=0, \pm 1/2, \pm 1, ...$, and
reflects on the fact that the soft part of the spectrum does not discriminate 
between bosonic or fermionic boundary conditions in the flux-free
case. The cooperon contribution may be rewritten using Poisson's 
resummation formula as
\be
p_{C} (t,\phi) =\frac V{2(4\pi D t)^2}
\sum_{[l]}e^{-2m|t|-\frac{l_{\mu}^2 L^2}{4D|t|}}\,\,
	\cos{(4\pi l_{\mu} \phi_{\mu})}
\label{pois}
\ee
with integer $l$'s. This result is in agreement with the one derived
by Montambaux~\cite{MONTAM} in disordered metals in lower dimensions.
The flux-accumulation in the cooperon part implies 
changes in the spectral correlations of the light quarks 
as we now show.

\vskip 0.3cm

{\bf 3.} To describe the spectral correlations associated with (\ref{01}) 
we will use semi-classical arguments for the two-point correlation
function $R(s, \phi)$ of the density of
eigenvalues~\cite{USPRL,MONTAM,IMRY}.
 Its spectral 
form factor $K(t,\phi )$ is defined as
\be
R(s , \phi) =\int_{-\infty}^{+\infty} dt\, e^{is\Delta t}\, K(t, \phi)
\label{spectral2}
\ee
where $\Delta=1/\rho V$ is a typical quantum spacing at zero virtuality 
and in the absence of a flux. For diffusive quarks in two-color QCD, 
$K(t,\phi )$ relates to the return probability through~\cite{USPRL,MONTAM,IMRY}
\be
K (t, \phi ) \approx\frac {\Delta^2 |t|}{(2\pi)^2} \, p(t, \phi )
\label{spectral1}
\ee
which is periodic with periodicity $\phi=0, \pm 1/2, \pm 1, ...$.

For $\phi\ll 1$, the flux periodicity may be neglected,
and for times larger than the ergodic time $\tau_{\rm erg}=L^2/D$ but
smaller than the Heisenberg time $t_H=1/\Delta$, the dominant contribution 
to the return probability stems from the zero modes in (\ref{des6}). Hence
\be
p(t, \phi) \approx \frac 12 e^{-2m |t|} \,\,\left( 1+ e^{-\tilde{F}^2 |t/t_H|}\right)
\label{small}
\ee
with
\be
\tilde{F}^2= 16\pi^2 \,\left(\frac {D}{L^2\Delta}\right) 
\, \phi_{\mu}\phi_{\mu}
=16\pi^2 \sigma_L\,\,\phi_{\mu}\phi_{\mu} \,.
\label{small1}
\ee
Here $\sigma_L=D/(\Delta L^2)=2(FL)^2/\pi $ is the dimensionless Ohmic 
conductance~\cite{USPRL}. In this limit, both the return probability
and the spectral form factor are only a function of the combination
$\sigma_L\phi^2$. The result~(\ref{small}) could be qualitatively derived
by noting that if $2\times 2\pi \phi$ is the typical accumulated flux
per winding path~\cite{NOTE3} in the quark return probability, then $l$
windings 
generate an accumulation $\Phi=4\pi l\phi$. Since the paths are diffusive
with an arc length $s=Ll\sim\sqrt{D t}$, then on the average 
\be
\la \Phi^2\ra =\la (4\pi l\phi)^2\ra \sim (4\pi \phi)^2 \frac {2Dt}{L^2}
	\,. 
\label{wind}
\ee
Hence 
\be
\la e^{i\Phi}\ra\sim e^{-\la \Phi^2\ra/2} = e^{-\tilde{F}^2 |t/t_H|}
\label{wind1}
\ee
which is the  dephasing appearing in (\ref{small}) between the 
diagonal and interference terms. An average $2\pi$ accumulation in 
$\Phi$ results from quark trajectories with a long proper time $t\sim 1/E_c$
where $E_c=D/L^2$ is the Thouless energy~\cite{THOULESS}. 

The spectral rigidity around zero virtuality (and also in bulk)
follows from~\cite{USPRL,EFETOV}
\be
\Sigma_2 (N, \phi) = \int_{-N}^{+N} ds\, (N -|s|) \, R(s, \phi)
\label{spec6}
\ee
with $N=E/\Delta\gg 1$. In particular
\be
\Sigma_2 (N, \phi) = 
\frac 1{2\pi^2} {\rm ln}\left(\left( 1+\frac {N^2}{\alpha^2}\right)
\left( 1+\frac {N^2}{\tilde{\alpha}^2}\right)\right)
\label{var}
\ee
with $\alpha=2m/\Delta$ and $\tilde{\alpha}=\alpha+\tilde{F}^2$. For
$\tilde{\alpha}\gg 1$ the spectral rigidity is reduced by about a factor of 2.
It is bracketed by the orthogonal (no flux) and unitary (finite flux) results. 

For large fluxes, the non-zero momentum contributions have to be retained
to enforce the proper $\phi=0, \pm 1/2, \pm 1, ...$
flux-periodicity. For simplicity, consider the 
case $\phi_{\mu}=(0,0,0,\phi)$ with only one-flux retained. For long proper
times, the zero modes along the $1,2,3$ directions contribute only, giving
\be
&&p(t, \phi) =\frac 12 e^{-2m |t|}\,\times \nonumber\\
&&\sum_n\left( e^{-4\pi^2 n^2 \sigma_L |t/t_H| } + 
e^{-4\pi^2 (n+2\phi)^2\sigma_L |t/t_H|}\right)
\label{per}
\ee
which is the analogue of a diffusion in d=1. For $t<\tau_{\rm erg}=L^2/D$ the 
diffusive paths are short and do not accumulate enough flux. The spectral 
rigidity in this case is just
\be
\Sigma_2 (N, \phi) = 
\frac 1{2\pi^2}\sum_n {\rm ln}\left[\left( 1+\frac {N^2}{\alpha_n^2}\right)
\left( 1+\frac {N^2}{\tilde{\alpha}_n^2}\right)\right]
\label{varx}
\ee
with $\alpha_n=\alpha+4\pi^2\sigma_L n^2$ and
$\tilde{\alpha}_n=\alpha+4\pi^2\sigma_L (n+2\phi)^2$.
For $N,\sigma_L\gg \alpha$, (\ref{varx}) simplifies to
\be
\Sigma_2 (N, \phi) =\Sigma_2 (N, 0) -\frac 1{\pi^2}{\rm ln}
\left(1+4\frac {\sigma_L}{\alpha}\, {\rm sin}^2 (2\pi \phi) \right)
\label{large}
\ee
in agreement with a result derived by Montambaux~\cite{MONTAM}
in the context of disordered metals. We note that for a small
flux (\ref{large}) is in agreement with (\ref{var}).
Interestingly enough, the conductance $\sigma_L$ of the chiral vacuum is
directly accessible from the spectral rigidity through (\ref{large})
providing for a direct measurement of this important quantity
in disordered QCD. (\ref{large}) may be assessed using current lattice 
QCD simulations.

\vskip 0.3cm
{\bf 4.}
Since the quark return probability and the spectral rigidity are 
sensitive to flux-variations in a finite Euclidean volume, they
could be  used to probe the flux-content of the two-color QCD vacuum
-- in particular monopole-antimonopole rich vacua which are likely
to generate flux-disordered vacua. Although our analysis so far has 
focused on Abelian electromagnetic Aharonov-Bohm fluxes, we speculate
that in the maximally projected gauge, colored magnetic monopoles and 
antimonopoles generate colored and Abelian fluxes that act randomly
on the light quarks in the vacuum.

If we consider an Abelian flux-disordered vacuum characterized by a 
Gaussian distributed flux with a mean 
\be
\la\la\phi_\mu\phi_\nu\ra\ra=\kappa_\mu^2\delta_{\mu\nu}
\label{measure}
\ee
then the quark return
probability can be easily estimated from (\ref{des7}) using the Poisson form
(\ref{pois}) for the cooperon part. If we split the quark return
probability $p =p_D+p_C$, then the diffuson part $p_D$ is flux-insensitive 
\be
\la\la p_D (t, \phi)\ra\ra =\frac V{2(4\pi D t)^2}\,\,e^{-2m|t|}
\label{vac1}
\ee
while the cooperon part $p_C$ is flux-sensitive 
\be
\la\la p_{C} (t,\phi)\ra\ra =\frac V{2(4\pi D t)^2}
\sum_{[l]}e^{-2m|t|-\frac{l_{\mu}^2 L^2}{4D|t|}
	-8\pi^2 \kappa_{\mu}^2 l_{\mu}^2}.
\label{vac2}
\ee
We note that the periodicity in
$\phi=0, \pm 1/2, \pm 1, ...$ of the quark return probability 
implies that the latter is likely insensitive to a $Z_2$ 
flux-disordered vacuum. If these effects extend to an Abelian flux-rich
vacuum, a simple way to detect them is to measure the relative ratio of the 
quark return probabilities for $N_c=2$ and $N_c=3$. A flux sensitivity 
implies a t-dependent ratio close to $1/2$ (as opposed to 1), assuming that 
the vacuum structure does not change appreciably from two to three colors.

\vskip 0.3cm
{\bf 5.}
Finally, since the transition from an orthogonal to unitary ensemble 
sets in the ergodic regime with $t>\tau_{\rm erg}$, the ensuing spectral 
statistics are amenable  to an analysis in 0-dimension (matrix model). The 
simplest realization that is chiral and embodies the essentials of a single 
flux  is \cite{NOTE}
\be
{\cal M}(\varphi) = \arr{im}{{\bf A}+i\varphi\, {\bf B}/\sqrt{N}}{{\bf A}
+i\varphi\,{\bf B}
/\sqrt{N}}{im}
\label{m2}
\ee
where ${\bf A}$ and ${\bf B}$ are real  $N\times N$ symmetric and 
antisymmetric random matrices respectively, with a fixed variance
$\Sigma^2/N$. For $\varphi=0$ and $m=0$ the matrices are chiral orthogonal, 
while for $\varphi\neq 0$ they are in general chiral unitary, hence the 
interpolation. The $1/\sqrt{N}$ in the off-diagonal matrix 
elements follows from second order perturbation theory since the typical
level shift induced by such terms is 
\eq
\Delta E = 
\Big\la \sum_n \f{|\Psi^*_n(I) {\bf B}_{IJ} \Psi_0(J)|^2}{E_n-E_0}\Big\ra_{{\bf A,B}}
\cdot \left(\f{\varphi}{\sqrt{N}}\right)^2
\eqx
where the $\Psi_n$'s are generic eigenfunctions of (\ref{m2}) with eigenvalues 
$E_n$ for ${\bf B}=0$. In the mesoscopic (ergodic) limit, the
sum is dominated by the modes where the energy denominator is of order 
$\Delta\sim 1/N$. If the states are delocalized then the eigenfunctions 
are of order $1/\sqrt{N}$. Since the sum over $I,J$ is of order $N^2$,
and the averaging over ${\bf B}$ is of order $\Sigma^2/N$, the second order
shift is $\Delta E\sim 1/N$ and comparable to the mesoscopic level
spacing $\Delta\sim 1/N$. This result is 
consistent with the observation made using non-chiral matrices~\cite{PANDEY}. 
We recall that in $d\neq 0$ dimension, the corresponding 
terms are down by $1/\sqrt[d]{V}$ as suggested by (\ref{001}).

\vskip 0.3cm
{\bf 6.}
We have shown that in two-color QCD, Abelian Aharonov-Bohm fluxes cause
changes in the quark return probability with important consequences on
the spectral form factor and the spectral statistics. The change is 
controlled by a universal function of $n^2\sigma_L\phi^2$, where $n$ is 
the number of applied fluxes. This analysis offers a simple measurement 
of the Ohmic conductance $\sigma_L$. When combined with the value of the
chiral condensate $\Sigma$, this leads to an independent assessment of the 
pion  weak-decay constant $F$  and a simple vindication of the GOR 
relation in two-color QCD.

The sensitivity of the quark return probability to Abelian fluxes for two-color
QCD raises  the interesting possibility of  addressing the nature of the 
confining fields in the QCD vacuum. Indeed, the quark return probability
for two colors can discriminate between a $Z_2$ flux-disordered phase and an
Abelian flux-disordered one, as it is blind to the former. Also, if the
flux-disordering is generic in two and three colors QCD, quantitative changes
in the proper time behavior of the pertinent quark return 
probabilities and spectral rigidities are expected. The present results can be 
tested by using lattice QCD simulations, or continuum models of the QCD 
vacuum~\cite{BOOK,DOSCH,SCHAEFER,MONOPOLE,PIERRE}.

\vskip 0.5cm
{\bf Acknowledgments}
\\

IZ would like to thank Edward Shuryak for a discussion.
This work was supported in part by the US DOE grant DE-FG-88ER40388, by the 
Polish Government Project (KBN) grants 2P03B04412 and 2P03B00814 and by the 
Hungarian grants FKFP-0126/1997 and OTKA-F026622. RJ was supported by the 
Fundation for Polish Science.


\begin{thebibliography}{99}
\bibitem{NOTE5}
Since the quark fields anticommute and $C^T=-C$, only the 
$\tau^2$ scalar diquark source is nonzero, implying  one
baryon (antibaryon) state for two flavors. For more
flavors we refer to~\cite{PESKIN}.
\bibitem{NOTE3}
The extra factor of 2 is due to the pair of quark paths.
\bibitem{NOTE}
The relation between the parameter $\varphi$ and the physical
flux $\phi$ follows intuitively from $\varphi^2 \equiv 
\sigma_L (2\pi \phi)^2$, since $\sigma_L$
sets the range of the matrix model~\cite{MONTAM}.
\end{thebibliography}
\end{document}